\documentclass[color]{aib}
\pdfoutput=1

\usepackage{makeidx}
\usepackage{subfigure}

\usepackage{graphicx}
\usepackage{paralist}

\AIBtitle{{\rwthblue SensorCloud: Towards the Interdisciplinary Development of a Trustworthy Platform for Globally Interconnected Sensors and Actuators}}
\AIBauthors{Michael Eggert, Roger H\"au\ss{}ling, Martin Henze, Lars Hermerschmidt, Ren\'e Hummen, Daniel Kerpen, Antonio Navarro P\'erez, Bernhard Rumpe, Dirk Thi{\ss}en, Klaus Wehrle}
\AIBnumber{2013}{13}
\AIBdate{October 2013}
\begin{document}

\makeAIBtitle

\title{SensorCloud: Towards the Interdisciplinary Development of a Trustworthy Platform for Globally Interconnected Sensors and Actuators\thanks{This paper has been presented as part of the workshop \textit{Wissenschaftliche Ergebnisse der Trusted Cloud Initiative} which took place in Karlsruhe, Germany on July 9-10, 2013.}}

\author{Michael Eggert\inst{2} \and Roger H\"au\ss{}ling\inst{2} \and Martin Henze\inst{1} \and Lars Hermerschmidt\inst{3} \and \\
Ren\'e Hummen\inst{1} \and Daniel Kerpen\inst{2} \and Antonio Navarro P\'erez\inst{3} \and Bernhard Rumpe\inst{3} \and Dirk Thi{\ss}en\inst{1} \and Klaus Wehrle\inst{1}}

\institute{
Communication and Distributed Systems, RWTH Aachen University, Germany\\
\texttt{\{henze,hummen,thissen,wehrle\}@comsys.rwth-aachen.de}
\and
Sociology of Technology and Organization, RWTH Aachen University, Germany\\
\texttt{\{meggert,rhaeussling,dkerpen\}@soziologie.rwth-aachen.de}
\and
Software Engineering, RWTH Aachen University, Germany\\
\texttt{\{hermerschmidt,perez,rumpe\}@se-rwth.de}
}

\maketitle

\begin{abstract}
Although Cloud Computing promises to lower IT costs and increase users' productivity in everyday life, the unattractive aspect of this new technology is that the user no longer owns all the devices which process personal data. 
To lower scepticism, the project SensorCloud investigates techniques to understand and compensate these adoption barriers in a scenario consisting of cloud applications that utilize sensors and actuators placed in private places.
This work provides an interdisciplinary overview of the social and technical core research challenges for the trustworthy integration of sensor and actuator devices with the Cloud Computing paradigm.
Most importantly, these challenges include
\begin{inparaenum}[i)]
\item ease of development,
\item security and privacy, and 
\item social dimensions of a cloud-based system which integrates into private life.
\end{inparaenum}
When these challenges are tackled in the development of future cloud systems, the attractiveness of new use cases in a sensor-enabled world will considerably be increased for users who currently do not trust the Cloud.
\end{abstract}

\section{Introduction}
Recent advances in human/nonhuman interaction networks, so-called Cyber-Physical (Social) Systems (CPS) which integrate computational, physical, and social processes \cite{GRSS12,L08}, continue to blur the boundaries between the physical and digital world \cite{hummen_sensorcloud_2012,henze_sensorcloud_2013} for a vast range of applications like health care, mobility, infrastructure, and manufacturing \cite{LYWZM11}.
These CPS often generate a huge amount of data that has to be stored and processed.
Additionally, it is desirable to aggregate data from multiple sources in order to generate higher value information.

The central approach of Cloud Computing with seemingly infinite computing and storage resources that can be scaled elastically has the potential to meet the requirements of CPS \cite{hummen_sensorcloud_2012,li_iot-cloud_2013}.
From the service developer's perspective, this ease of use is largely to be attributed to the separation of responsibilities and the flexible, on-demand billing model introduced with Cloud Computing.
More precisely, one or more providers are made responsible for maintaining and scaling the central computing and network infrastructure to the demand of their resource consumers. 
As a result, service developers are unburdened from managing individual compute units and the reliable interconnection between each other as well as with the Internet.
However, to fully utilize the Cloud Computing paradigm, distributed programming models must be employed in developing a service.
This renders the development of clouds a challenging task and demands for further development tool support.

Furthermore, the involvement of multiple providers in the provisioning of clouds opens new vectors of attacks against potentially sensitive data that is stored or transmitted within the cloud environment.
As such, Cloud Computing stands in stark contrast to previous IT outsourcing activities that involve a single clearly defined entity offering its IT services.
Cloud Computing, however, inherently implies multi-tenancy of the offered system.
Each tenant (e.g., a service provider or an administrator of the cloud infrastructure provider) may be interested in another tenant's data.
This makes Cloud Computing highly challenging from the security perspective.

Finally, clouds are commonly consumed by non-expert users.
Hence, the presentation of data as well as the underlying security model must be intuitive and understandable by the common user.
Otherwise, adoption barriers prevent the usage of a technically sound, highly secure, but unusable system.
Thus, user acceptance testing is a major concern for massive-scale services.

This work is structured as follows: After this introduction, we first present the vision and scenario of the SensorCloud project.
Afterwards, we identify and discuss research challenges when outsourcing storage and processing of sensor data to the cloud.
We identify three core research challenges:
\begin{inparaenum}[i)]
\item ease of development,
\item security and privacy, and 
\item social dimensions of a cloud based system which integrates into our private life.
\end{inparaenum}
Thus, we focus on these challenges and discuss approaches to address them.
These approaches will help to develop cloud solutions that users can confide there sensitive sensor data.

\section{SensorCloud Vision and Scenario}

The SensorCloud project, incorporating partners from industry and academia, is a member of the project cluster Trusted Cloud which is funded by the German Federal Ministry of Economics and Technology. 
Trusted Cloud aims to develop and evaluate innovative, robust, secure and law conform Cloud Computing solutions. 
Pilot projects will apply these solutions to industrial demands and thus demonstrate the benefits and trustworthiness of Cloud Computing. 
These pilots will especially address the current skepticism towards Cloud Computing observable in small and medium enterprises and their consumers.

The goal of SensorCloud is to develop a cloud-based platform that integrates basically every kind of distributed internet-connected sensor and actuator devices. 
Such devices may be located at a variety of places like personal homes, offices, cars, industrial plants, or outdoor areas.
As scarcity of limited resources gets an ever growing issue, the controlled use of these resources gets a success factor for modern industry and future smart homes. 
In the SensorCloud project, we use the smart home scenario to capture requirements of home users and to validate assumptions necessary when designing technical solutions.
The smart home scenario foresees a not to far future, where all devices, handles, and general interaction points residents or visitors have with a house are equipped with a sensor resp. actuator. 
A house with sensors and actuators is not smart in itself but enables an IT system to have a rich interaction with the users of the house. 
In order to make the house smart, SensorCloud serves as a central hub to which such devices can be connected. 
It aggregates their data and controls their functions. 
Thereby, it manages devices and data from different domains, places and owners at the same time.
On top of its base of devices and data, SensorCloud provides an extensible and flexible service platform that is populated by ``apps'' from third-party developers. 
These apps leverage the potential of the centrally aggregated devices and data, providing a rich, diverse and innovative supply of end user functionality. 
Bringing together owners and suppliers of internet-connected sensors and actuators, developers of services working on these devices, and the provider of the integrating cloud platform, the SensorCloud project enables a marketplace for innovative customer solutions that minimizes costs and hurdles for all involved players.

\section{Research Challenges}
\label{challenges}

We identify three major challenges when integrating storage and processing of sensor data with the cloud paradigm:
\begin{inparaenum}[i)]
\item development tool support is needed in order to assist the developer in implementing efficient, distributed services that can handle large amounts of data,
\item security and privacy mechanisms have to be considered as first-class citizens in the system design in order to cater to the security requirements of the inherently multi-tenant cloud system, and
\item acceptance of the end-user has to be assured in order to overcome potential acceptance barriers of a secure cloud system.
\end{inparaenum}
We now further detail the respective challenges.

\subsubsection{Ease of Development.}
Cloud Computing unites approaches and technologies from different areas \cite{LLL09,SSK10,RDA09}:
The management of the physical cloud infrastructure in an infrastructure layer often referred to as Infrastructure as a Service (IaaS) is the most mature and most commonly used cloud technology. Virtualization and dynamic resource allocation are the main technologies in this area. From a software engineering point of view the specification of the infrastructure resources used by an application as well as the quality attributes and costs of these resources are of special interest \cite{LLL09}.
On top of this layer a platform layer called Platform as a Service (PaaS) is located which provides common functionality. This layer is typically realized with component-based middleware which utilizes the Service Oriented Architecture (SOA) pattern.

Although Cloud Computing facilitates scalable and cheap operations, it adds additional complexity to the design and implementation of services based on such a cloud. 
The service's software has to be designed appropriately to deliver the desired cloud properties of massive scaling and parallelization over heterogeneous computation and networking resources. 
Thereby, cloud specific software engineering faces known challenges at a larger scale as well as new challenges.

In order to successfully design, develop, deploy, and operate Cloud Computing services for specific tasks, a service developer needs diverse skills, ranging from understanding the behavior and background of cloud service users, over concepts and architectures to ensure security and privacy, up to the correct and scalable implementation of such concepts which can run for years in the cloud. This is a non-trivial and error prone task. Therefore, approaches are needed to simplify the tasks of the developer. For instance, in model-driven software engineering \cite{GS03} models are used to abstract from technical details (e.g., how to build a system) and to focus on higher level requirements from the stakeholders. So, models are used to capture the requirements in a more natural, non-technical way. Several modeling languages or domain specific languages (DSL) have been introduced for this purpose (e.g., \cite{DKV00}), each focusing on a specific problem domain. The models are then used to automatically or semi-automatically generate the final technical system. To do so, they need to have a precise semantic.

The PaaS lacks mature development approaches which cope with cloud specific aspects. Especially the question of proper modeling of and programming languages for cloud systems is an open question. Only specialized solutions, e.g., MapReduce \cite{DG08}, are widely used. cloud specific agile methods are an open topic as well. Especially the question of efficient testing of service platforms and services is important to answer in order to adopt agile processes to cloud service development \cite{RTS10}.

\subsubsection{Security and Privacy.}
Data sensed by CPSs often contains sensitive information~\cite{hummen_sensorcloud_2012}.
This is not only restricted to the sensed raw data but also applies to meta-information, e.g., time or location.
Thus, the owner of the data often does not want to reveal her data unconditionally to others.
Instead, she may strive to share her sensitive information only with a few, carefully selected cloud services, while concealing her data from the cloud provider or other cloud services that she does not fully trust.
Hence, the cloud provider must be able to guarantee the confidentiality and protection of data while being stored and processed by the data owner-defined set of clouds.
Otherwise, the data owner may completely disregard the option of storing and processing her sensor data in the cloud.
The root cause of this adoption barrier, as previously identified in~\cite{chow_controlling_2009,henze_cloud-annotations_2013,pearson_cloud-issues_2010,henze_cloud-data-handling_2013}, is the fact that the \emph{data owner loses control over her data} in the cloud.

Typical state-of-the-art approaches to securing storage and processing of data in the cloud aim at providing hard security guarantees.
For this purpose, they leverage technologies such as (fully) homomorphic encryption or trusted platform modules~\cite{santos_trusted-cloud_2009,gentry_fhe_2010,wallom_mytrustedcloud_2011}.
We argue that these approaches are not sufficient for storing and processing sensitive sensor data in the cloud.
These approaches either lead to an excessive encryption overhead when applied to rather small (only a few bytes) sensor readings~\cite{danezis_towards_2011} or do not offer the necessary control of users over their outsourced data.
Additionally, it has been shown that encryption alone is not sufficient to guarantee privacy in Cloud Computing~\cite{van_dijk_impossibility_2010}.

Thus, a core research challenge when outsourcing the storage and processing of possibly sensitive sensor data to the cloud is the design of a \emph{practically viable security architecture}.
This security architecture has to enable the data owner to stay in control over her data.
To this end, it has to offer transparent measures to protect data already in the sensor network, to grant fine-grained access to user-selected cloud services, and to isolate cloud services at the platform level.

\subsubsection{The Socio-technical Condition.}
The Cloud Computing paradigm constitutes a relatively new approach to information and communication technology. Its qualities, from a user perspective, differ greatly from hitherto well-known and institutionalized ways to handle data and computer technology. Based on the idea of releasing the storage and processing of individual data to the -- spatially and organizationally distributed -- cloud, the user is confronted with a significantly higher degree of complexity of the system used. Thus, the task of developing a potentially successful cloud architecture exceeds the mere fulfillment of well-defined technological demands. Instead, it should be understood as the challenge to materialize the vision of a certain socio-technical system, especially in the case of a technology like SensorCloud which will serve as a platform for the connection of a multitude of sensor and actuator equipped devices constantly analyzing and influencing even our physical everyday environment.

Constraints to implementation should be addressed as early as possible in the course of the development process. Hence, one of our major points is that, besides all technical feasibilities, the potential success of innovative cloud-based systems and services needs to be explored in the context of particular social dimensions such as user acceptance, usability, and environmental conditions already in the early stages of development.

Unfortunately, to our knowledge only few related sociological work has currently addressed Cloud Computing related issues. Though recent reviews refer to some cloud-related contributions from a social science perspective \cite{ven12,yan12}, particularly sociology has so far only rudimentarily dealt with the topic. Nevertheless, a holistic analysis of socio-technical constellations and relationships is especially needed in our context of information and communication technologies, because such a perspective allows to take human and non-human (inter-)action and (inter-)activity into account \cite{rh12}. Investigating on the technology's acceptability and chances for its contextualization is therefore decisive for the success of the development project.

Assuming major developments in human-technology relationships implicated by the development of cloud technologies, indeed, it seems not sufficient to simply identify drivers and barriers for the diffusion of respective technologies and services. Moreover, we seek to sketch a concept of a human-cloud interaction \cite{dahm06,rh09,hei04} which is embedded in social practices and social structures \cite{cerulo09,ram09} in order to understand and support future developments in the area of Cloud Computing. This, however, cannot be realized without giving regard to the processes and actors which are responsible for the future of Cloud Computing themselves. Hence, an encompassing analysis of the SensorCloud innovation process completes the bundle of research challenges posed to our sociological interest by the technological project.

\section{Development of Cloud Software}

Up to now we discussed challenges which have to be solved when developing a cloud. As discussed the development of cloud service platforms and services implemented on top of these platforms is a software engineering challenge itself.

\begin{figure}[t]
\centering
\includegraphics[width=0.9\textwidth]{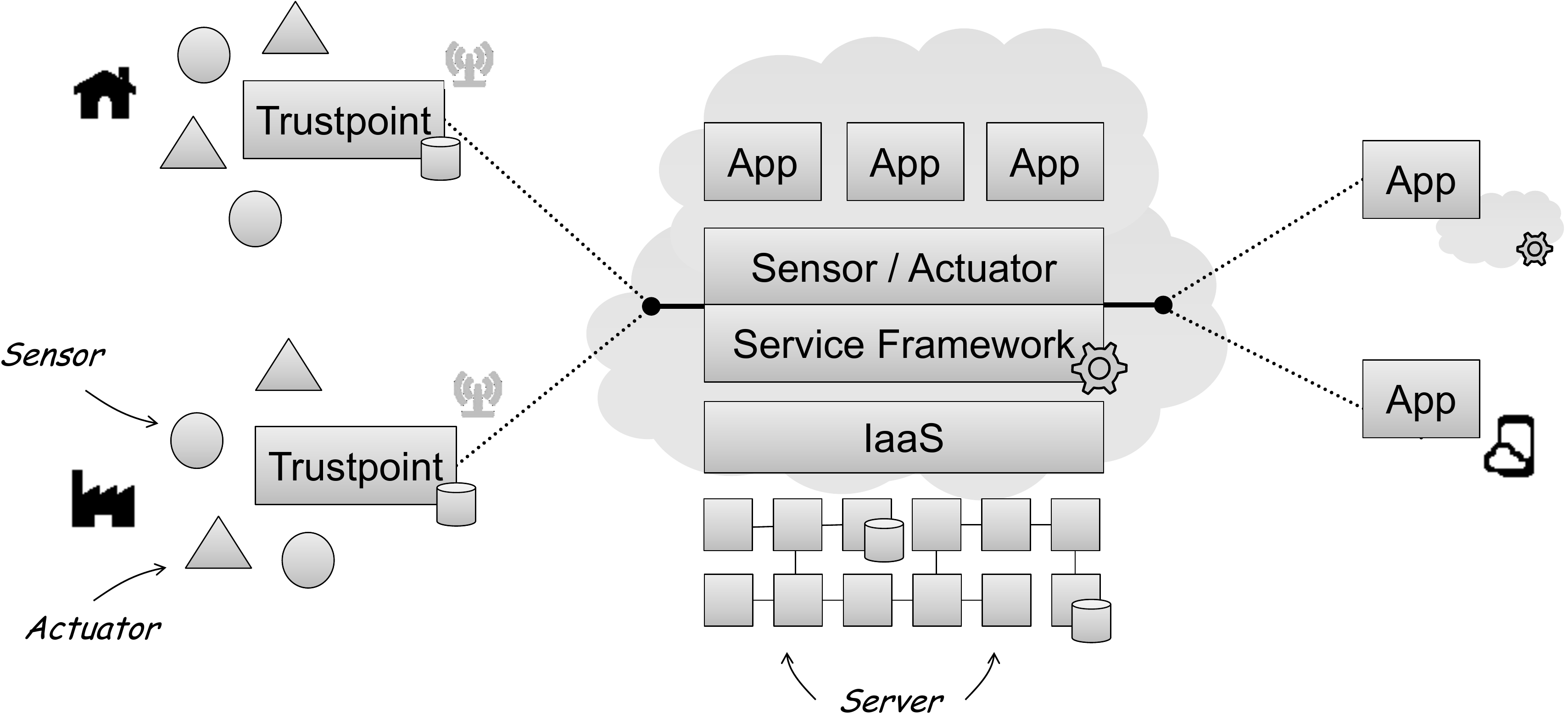}
\caption{Overview of the cloud architecture}
\label{fig:se-cloud-layer}
\end{figure}

A cloud service platform is a generalized execution environment for software instances, which implement a particular functionality and share technical and/or business domain specific similarities. Software instances may be called applications, components, services, and so forth, but we refer to all of them as \emph{apps}. The cloud service platform supports developers in developing apps by providing the commonalities as ready-made functional building blocks which can be accessed over simple and clearly defined interfaces. At run-time, the cloud service platform provides the apps with infrastructural resources (processing power, data persistence, ...).
As part of the project SensorCloud we developed a multilayered architecture of cloud service platforms which utilizes the Cloud Computing paradigm. Figure \ref{fig:se-cloud-layer} gives an overview of the layered cloud architectures. The cloud architecture consists of four layers: The \emph{Infrastructure as a Service} layer (IaaS), the \emph{Cloud Service Framework} layer, the \emph{Sensor/Actuator} layer and the \emph{Apps} layer.
The \emph{Sensor/Actuator} layer uses the cloud-specific basic functionality provided by the Cloud Service Framework and adds sensor and actuator domain specific reusable functions, which are part of several use cases. An app developer thus can concentrate on the use cases which are specific to her app. 
The development of apps which use the cloud service platform and the development of the platform itself are performed in two separate development processes by distinct groups of developers.
To improve the development of the cloud service platform and the services, we provide both development teams with software engineering methods, concepts, and tools. This support is focused on the software engineering principals agile methods, model-driven software engineering, and component-based software architecture.

\subsection{Agile Software Development}
The fundamental principal of agile software development is to set inflexible, bureaucratic processes with long-running up front planed development steps aside. Instead, the software development is broken down to activities, which are performed in short iterative cycles. At the end of each cycle the result of each activity is reflected and the next cycle is planed taken this reflection into account. The activities are guided by agile methods and tools, which aim to reduce organizational efforts and maximize the effort spend in constructive work. Every cycle starts by breaking down the current requirements to the final product down to activities for the starting cycle. The cycle ends with a reflection about the results, which may cause a change or addition of requirements. A central goal at the beginning of the project is to reach a state where an early version of the final software system is running. From this point on the software should be always in a state where it can be released containing all features added in the last cycle. This running system is then used to validate customer requirements as soon as possible against the actual software system. Changes in requirements resulting from this early validation can then be implemented with minimal effort.
Agile software development is best fitting for projects where the challenges are unclear and new requirements arise during the project and others change. The method enables a project team to validate requirements in short intervals and lowers the effort for changes resulting from changing requirements. Thereby agile methods are valuable especially for highly innovative research projects like SensorCloud where new technologies are build. Development decisions in these projects are mostly of explorative nature and typically have to be implemented with small effort and in different versions which then are compared and only one alternative may be accepted for future development.

\subsection{Model-Driven Software Engineering}
The foundation of model-driven software engineering is to represent the different aspects of the software system with suitable explicit, dedicated models which use abstraction to focus on the relevant aspects. The models are used as foundation for analytic, constructive, and communicative software engineering activities. The modeling language in which models are expressed geared to the mental and vocabulary of concepts of the problem domain. As models are more abstract and concise then source code of software they ease the understanding of the software to be developed and thereby analytical software engineering activities.

In the SensorCloud project we focus on constructive activities which aim to automatically derive a significant part of the final system implementation from models. One part of this is the code generation where the system implementation is partly or completely generated from models. The increased automation leads to enhanced quality in the resulting software. For example wide spread changes on a system, which require significant manual effort in conventional development can be realized by small changes on the model with less effort in model-driven development. As model-driven software engineering enhances development efficiency it supports agile methods and therefore benefits to SensorCloud as discussed before. In addition the development of complex systems where several complicate aspects interact like in SensorCloud benefits from the dedicated, compact and understandable representation of single aspects in specialized models. 

\subsection{Component-Based Software Engineering}
The fundamental basis for component-based software engineering is the fundamental software engineering principal ``separation of concerns''. The key to handle the development of complex systems therefore lies in the modularization of the system in small components, where each component fulfills a precisely defined task. The system is then constructed by combining these components. Components may contain other components which leads to a hierarchical decomposition of the system. Components implement the functionality of the system by collaborative interaction among each other. Components hide there internal structure and communicate with other components over channels with defined interfaces. This component network represents the overall architecture of the software, which abstracts from the implementation details. In addition, components in a distributed system  represent the atomic units which can be distributed. For cloud based architectures, aspects at the overall architectural level are of special relevance. These aspects include distribution of components, realization of the communication between components, elastic replication of components, fault tolerance of components, monitoring of components, as well as the logging of interactions between components.

In the project SensorCloud we study the combination of the principals agility, model-driven and component-based development in cloud based software in general and in the specific context of SensorCloud. Our approach integrates agile development principles with model-based development techniques. At its core is a set of cloud-specific architecture-oriented modeling languages. Those languages allow for the formal specification of the software architecture of cloud software as well as properties of that architecture like distribution, elasticity, robustness, and monitoring. Based on these languages, we develop a set of tools that allow for analysis/simulation, code generation, and testing.

\section{Practically Viable Security Architecture}

As identified in Section \ref{challenges}, one core research challenge that has to be addressed when integrating sensor data with the cloud is the design of a practically viable security architecture.
The main design goal of our security architecture is to enable the data owner to stay in control over her data when outsourcing storage and processing of sensor data to the cloud. 

\subsection{Trust Domains}

\begin{figure}[t]
\subfigure[Data flows through different trust domains.]{\label{fig:comsys-trust-domains}\includegraphics[width=0.48\textwidth]{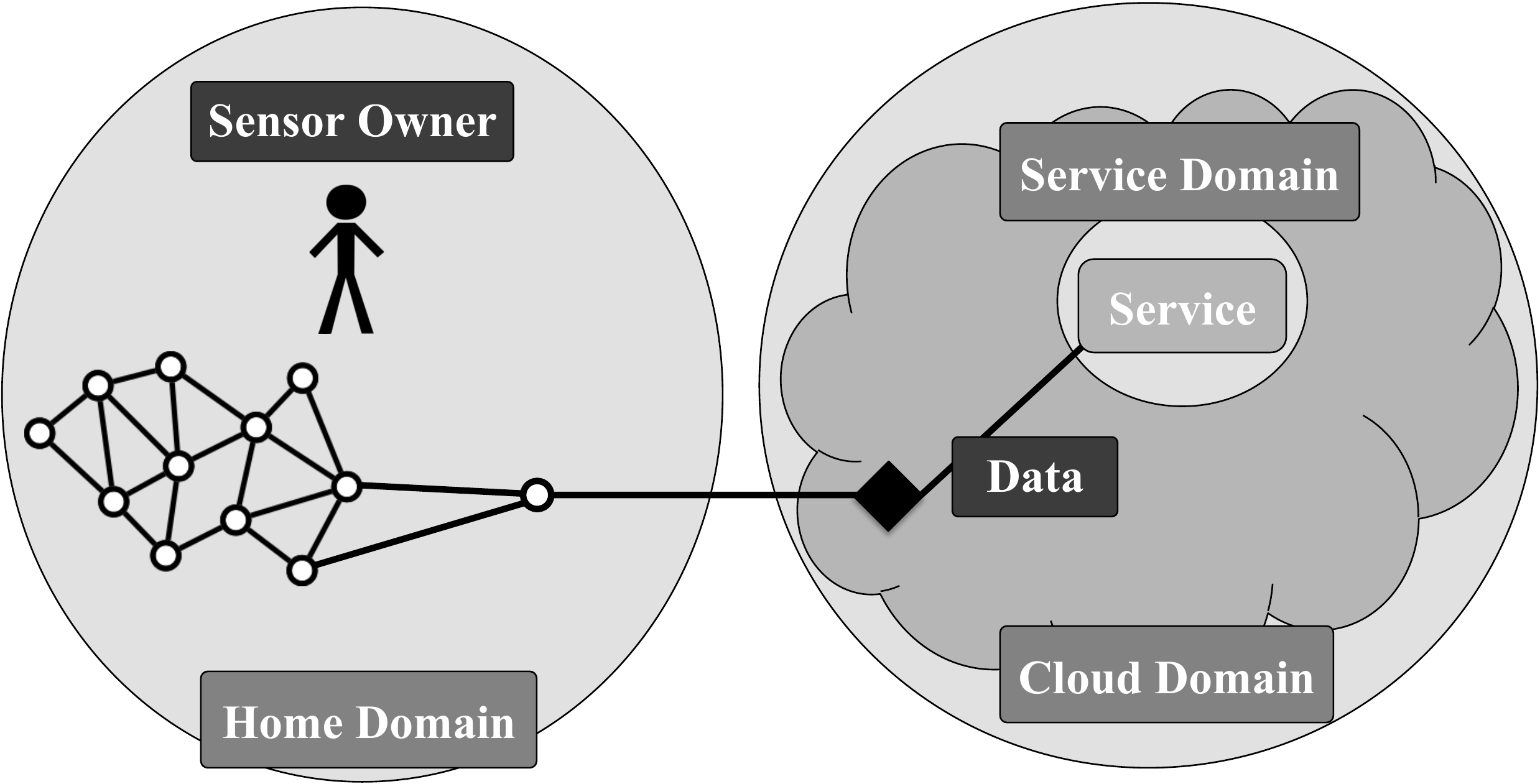}}
\hfill
\subfigure[The trust point applies transport security and object security.]{\label{fig:comsys-trust-point}\includegraphics[width=0.48\textwidth]{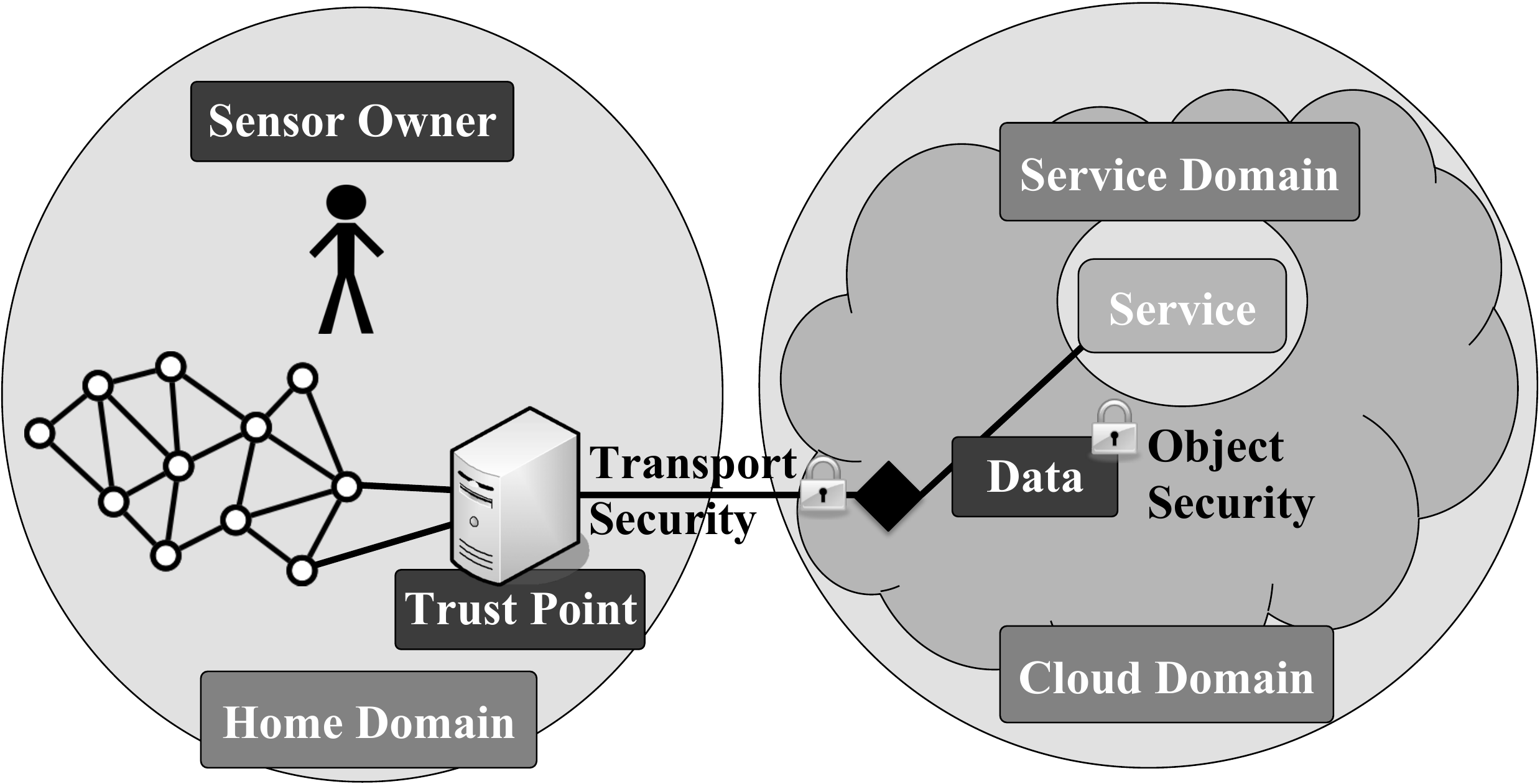}}
\label{fig:comsys-trust}
\caption{Data flowing from the home domain up to the service domain has to be secured before it leaves the home domain which is controlled by the sensor owner.}
\end{figure}

As a basis for the development of our security architecture, we identified three high-level trust domains that are essential for our security architecture: the home domain, the cloud domain, and the service domain.
Figure \ref{fig:comsys-trust-domains} gives an overview of these domains and depicts their interaction.
We now discuss these trust domains and the underlying assumptions in more detail.

The \emph{home domain} of a sensor owner consists of a multitude of sensor and actuator devices that are interconnected in a local sensor network.
The sensor network is connected to the internet via a gateway located at the border of the home domain.
We assume that only authorized devices can participate in the sensor network and that third parties cannot overhear communication within the home domain.
This can, e.g., be achieved using wired connections or secure wireless communication.
Thus, the sensor owner can fully trust all devices that are located within her home domain.
Still, trusted communication ends as soon as the gateway passes data to entities outside the sensor network.
For the \emph{cloud domain}, the cloud provider has to guarantee trust into the cloud platform.
To this end, it has to take the necessary technical (e.g., cryptography), organizational (e.g., security policies), and legal (e.g., contracts) measures.
For both, data owner and service provider, the cloud domain is a trusted domain.
The \emph{service domain} consists of isolated service instances.
Each service instance is a trust domain for the respective sensor owner.
It is the task of the cloud provider to establish trust into the service domain using technical (e.g., virtualization), organizational (e.g., policies), and legal (e.g., contracts) measures.
The trust domain ends as soon as data is forwarded to a different entity and thus leaves the service instance.

Inside the respective trust domain, the responsibility to keep data available, confidential, and unmodified can be attributed to a single entity.
For example, a service provider is responsible for keeping data that the service may access undisclosed from other third parties.
However, data in transit is at risk.
To securely bridge between the identified trust domains, we propose to use a combination of transport and object-related security mechanisms.
Specifically, we employ transport security in order to protect data during transfer from the gateway to the cloud.
Moreover, our security architecture additionally protects data at rest (i.e., storage) and during transmission in the cloud by means of additional object-based security that is applied per data item.

\subsection{Trust-Point-Based Security Architecture}

When bridging trust domains, we aim to allow the data owner to stay in control over her data in the cloud.
Hence, there exist two main requirements when securely outsourcing data storage and processing to the cloud:
\begin{enumerate}
\item Storage and transmission of sensor data must be protected against illegitimate access. Most importantly, unauthorized services or cloud-external entities must not be able to access or modify sensor data unnoticeably.
\item Access to sensor data in the cloud requires explicit approval by the data owner. As a result, only data owner-selected services have access to sensor data, thus limiting permission to a small set of privileged services.
\end{enumerate}

From these requirements, we draw two conclusions. 
First, data has to be secured whenever it is stored or forwarded outside a trusted domain.
Second, the control over the data access has to take place in the control domain of the sensor owner, i.e., at her home domain.
Thus, we introduce the \emph{trust point} as a control instance at the (network) border of the home domain (see Figure \ref{fig:comsys-trust-point}).
The trust point maintains a secure connection between the home domain and the cloud domain.
Additionally, it manages access of service instances to sensor data of the respective sensor owner towards the cloud domain and the service domain.
The trust point thus acts as a representative of the sensor owner.
Introducing a trusted entity similar to our trust point has successfully been proposed for different scenarios in the past, e.g., in the context of Wi-Fi-sharing communities~\cite{heer_pisa-sa_2010} or for intelligent energy networks in Germany~\cite{pp_smart_meter}.

The communication between trust point and SensorCloud takes place authenticated and encrypted at all times (transport security).
In addition to the transport security measures, the trust point applies object security measures to the sensor data.
This allows for secure storage in the cloud PaaS and realizes the access control for service instances.
Before the trust point forwards sensor data to the cloud, it thus encrypts sensor data (and possibly meta data) according to the demand of the sensor owner and applies integrity protection mechanisms.
The permission to access data in the cloud can hence only be granted by the trust point.
For this, the trust point provides, upon request of the sensor owner, a service instance with the necessary keying material for decrypting the selected sensor data.
By introducing the trust point as a local control instance, we offer the sensor owner control and transparency over the access to her data even when they are stored and processed in the cloud.

\section{Socio-technical Innovations for the Cloud}

The following section outlines central foci of interest stemming from the perspective of sociological analysis of human-cloud interaction. They are derived from the already outlined sociological research challenges and present intermediate results of our current project.

The aim of the sociological work within the project SensorCloud may roughly be seen as comprising the following different streams of interest: On the one hand there are the prospective users of SensorCloud technologies, their estimations towards the technological propositions and thus the acceptability and consequently the acceptance of SensorCloud within a certain context. These aspects derive from the wider context reflecting Cloud Computing technologies' potential to change human-computer interactivity. On the other hand, it is the development process itself which is subject to our sociological interest stemming from a huge tradition of sociological innovation research \cite{dou06}.

\begin{figure}[t]
\centering
\includegraphics[width=0.8\textwidth]{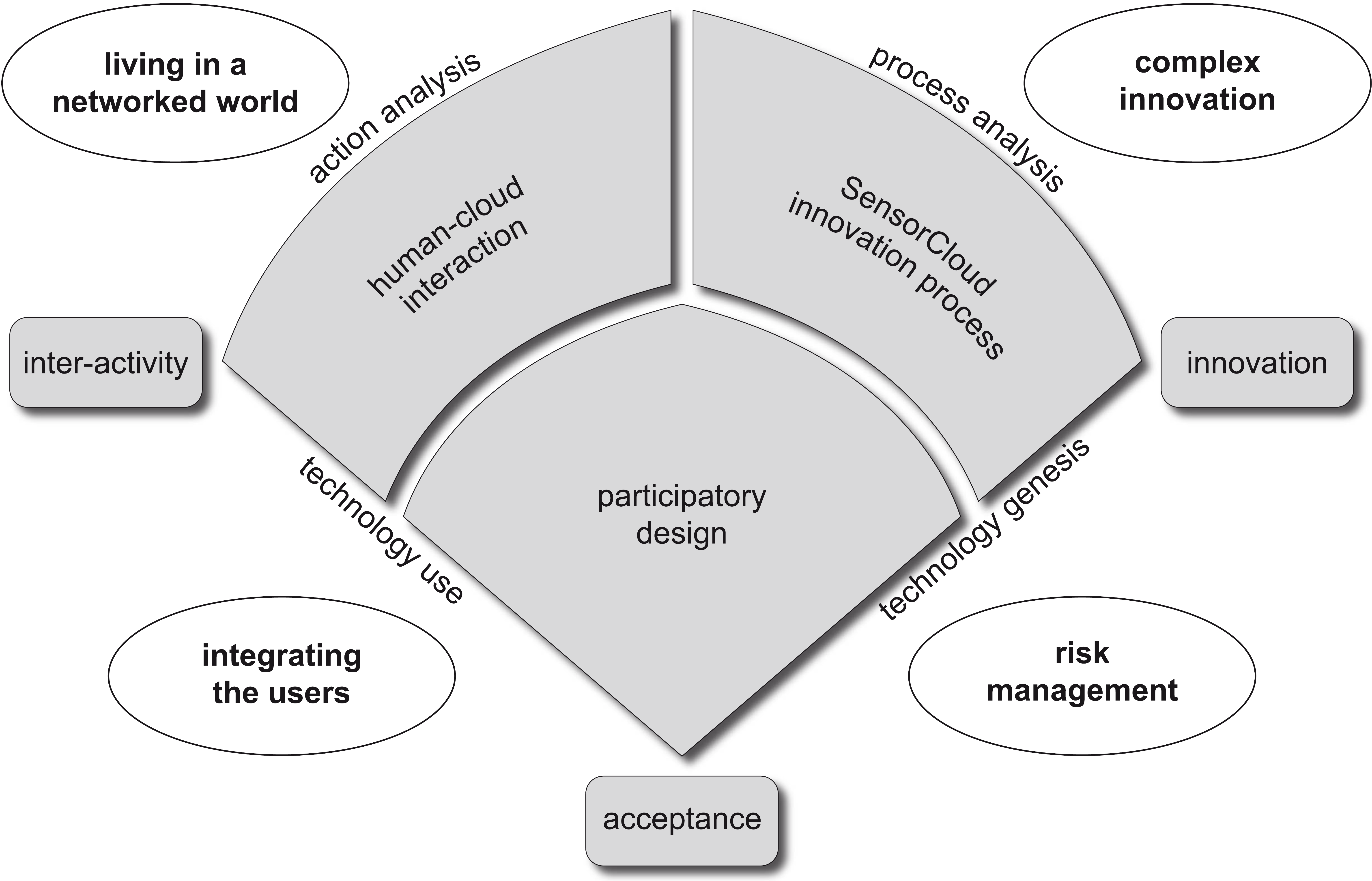}
\caption{Dimensions of sociological analysis}
\label{fig:tech-soz}
\end{figure}

Structured around the three poles of research on \begin{inparaenum}[(1)]\item interactivity between humans and technology, \item innovation processes, and \item social acceptance of emerging technologies\end{inparaenum}, these streams result in a design consisting of the research strategies sketched in Fig. \ref{fig:tech-soz} and outlined in the following sections. By pursuing this strategy, on the one hand, we seek to actively contribute to the success of the SensorCloud development project. We accomplish this by informing the partners about the prerequisites for acceptability of the technological system to be developed. On the other hand, we aim at gaining a comprehensive picture of the SensorCloud innovation process as a whole. From this efforts we expect us being able to draw general conclusions for future research on both the question of human-cloud interactivity and the emergence of innovations for Cloud Computing technologies and smart environments.

\subsection{Living in a Networked World}
Cloud Computing and ubiquitous networking of an ever increasing multiplicity of diverse artifacts undoubtedly change the way in which people deal with technology and implicate transformations in the human-object-relationship. Technology may no more be conceived as a mere tool to achieve certain goals, but increasingly becomes a kind of a partner in interaction in all kinds of contexts \cite{kor09}. At the same time, technology vanishes to the background and fulfills its task without being actively influenced by the users \cite{wei91}. Within the SensorCloud project, we observe these transformations and explore this novel way of dealing with technology by means of technographic field studies in order to identify and characterize this new quality in the relation between humans and their technological environments.
\subsection{Complex Innovational Setting}
Technological innovations do not arise in a vacuum, and the idea of a single genius inventor may consequently be regarded as obsolete, too. Instead, complex innovations like the SensorCloud project result from joint efforts of heterogeneous actor-networks \cite{rh07} from science, (corporate) research, political actors, and intermediaries, all of which try to pursue their own individual goals in developing and implementing such a technological project. Therefore, the sociological analysis of technology development critically accompanies the entire SensorCloud development process from its very beginning. It examines how the actors balance their respective interests in order to come to mutually shared understandings of process-related problems, decisions, and common solutions. Already from the launch of the project, such a sociological perspective of technology development may obtain evidence on the chances for innovation under the condition of conflicting political, legal, social and economic interests and requirements.
\subsection{Participatory Design}
One of the crucial elements for the success or failure of an innovation is its potential users. Especially in the context of a technology such as SensorCloud, whose benefits as a network effect good \cite{katz85} become more evident the larger its user base grows, it is therefore essential to integrate the users' perspective into the development process as early as possible. Against the background of a possible SensorCloud deployment in a smart home scenario, our interest focuses on users belonging to the so-called ``New Middle Class Milieu''\footnote{\url{http://www.sinus-institut.de/en}} with their hopes and expectations, fears and concerns. In terms of user-centered participatory design, qualitative and quantitative studies help us to identify drivers and barriers for the acceptance of the SensorCloud technology. In short, users are directly involved in the project and become an active element of technology development.
\subsection{Managing Risks}
In general, engineers and designers may be considered as the driving forces behind technology innovation projects. Therefore, observation of engineers' and designers' working practices, for example in the laboratories of their research institutes, should be given high priority. But their work is also often confronted with the risk of incurring concepts and ideas which may miss actual needs on the user side. Another problem may be the encounter of cultural, structural, and organizational barriers in the process of development and diffusion \cite{kno99}. Here, the sociology of technology can contribute significantly to the SensorCloud innovation process by addressing such uncertainty by continuously reflecting the results to the involved project partners. Furthermore, we help to spur the project's course by implementing joint future workshops of developers, users, and other relevant groups. In such workshops, the different stakeholders meet, share time discussing topics of technological innovation, and (more or less) mutually agree on and incorporate their different views, ideas, and proposals about the topic \cite{jun87,bell97}. However, such future workshops do not only function as arenas to thoroughly deliberate ideas and consequences of the project in question. In addition, such dialogue serves as the basis for developing a trustful relationship between the different stakeholders.

\section{Conclusion}

As Cloud Computing enables CPS scenarios such as SensorCloud, where potentially privacy relevant data is processed, the development of such systems faces technical and non technical challenges, as we stated in our work. Within this paper, we presented the three directions of research which integrate efforts to build trustworthy cloud platforms and services. From the socio-technical perspective, we presented a framework useful for the sociological analysis of human-cloud interaction, focusing on the interactivity between humans and technology, on innovation processes, and on the social acceptance of emerging technologies.
From the technical security and privacy perspective we showed a practical solution for users of cloud services to better control there data in the cloud. And from the software engineering perspective we showed that foundations like component-based and model-driven software engineering can be combined to support agile methods when developing cloud systems.
These approaches will help to develop trustworthy cloud services which users can confide there data from all kind of CPS.

\subsubsection{Acknowledgments.} The authors would like to thank all members of the SensorCloud consortium for the inspiring discussions on the concepts described in this paper.
A special thanks goes to all participants of the sociological acceptance study.
The SensorCloud project is funded by the German Federal Ministry of Economics and Technology under project funding reference number 01MD11049.
The responsibility for the content of this publication lies with the authors.

\cleardoublepage
\section*{Aachener Informatik-Berichte}
\newfont{\sss}{cmr10 scaled 1000}
\newfont{\bbb}{cmbx10 scaled 1000}
\sss

{\bbb This list contains all technical reports published
  during the past three years.
  A complete list of reports dating back to 1987 is available from:
\begin{center}
  \url{http://aib.informatik.rwth-aachen.de/}
\end{center}
  To obtain copies please consult the above URL or send your request
  to:
\begin{center}
  Informatik-Bibliothek, RWTH Aachen, Ahornstr.~55, 52056 Aachen,\\
  Email: \email{biblio@informatik.rwth-aachen.de }
\end{center}}\bigskip

\begin{longtable}{lp{11cm}}

2010-01 $^\ast$ &Fachgruppe Informatik:      Jahresbericht 2010\\
2010-02 & Daniel Neider, Christof L\"{o}ding:         Learning Visibly One-Counter Automata in Polynomial Time\\
2010-03 & Holger Krahn:         MontiCore: Agile Entwicklung von dom\"{a}nenspezifischen Sprachen im Software-Engineering\\
2010-04 & Ren\'{e} W\"{o}rzberger:         Management dynamischer Gesch\"{a}ftsprozesse auf Basis statischer Prozessmanagementsysteme\\
2010-05 & Daniel Retkowitz:         Softwareunterst\"{u}tzung f\"{u}r adaptive eHome-Systeme\\
2010-06 & Taolue Chen, Tingting Han, Joost-Pieter Katoen, Alexandru Mereacre:         Computing maximum reachability probabilities in Markovian timed automata\\
2010-07 & George B. Mertzios:         A New Intersection Model for Multitolerance Graphs, Hierarchy, and Efficient Algorithms\\
2010-08 & Carsten Otto, Marc Brockschmidt, Christian von Essen, J\"{u}rgen Giesl:         Automated Termination Analysis of Java Bytecode by Term Rewriting\\
2010-09 & George B. Mertzios, Shmuel Zaks:         The Structure of the Intersection of Tolerance and Cocomparability Graphs\\
2010-10 & Peter Schneider-Kamp, J\"{u}rgen Giesl, Thomas Str\"{o}der, Alexander Serebrenik, Ren\'{e} Thiemann:         Automated Termination Analysis for Logic Programs with Cut\\
2010-11 & Martin Zimmermann:         Parametric LTL Games\\
2010-12 & Thomas Str\"{o}der, Peter Schneider-Kamp, J\"{u}rgen Giesl:         Dependency Triples for Improving Termination Analysis of Logic Programs with Cut\\
2010-13 & Ashraf Armoush:         Design Patterns for Safety-Critical Embedded Systems\\
2010-14 & Michael Codish, Carsten Fuhs, J\"{u}rgen Giesl, Peter Schneider-Kamp:         Lazy Abstraction for Size-Change Termination\\
2010-15 & Marc Brockschmidt, Carsten Otto, Christian von Essen, J\"{u}rgen Giesl:         Termination Graphs for Java Bytecode\\
2010-16 & Christian Berger:         Automating Acceptance Tests for Sensor- and Actuator-based Systems on the Example of Autonomous Vehicles\\
2010-17 & Hans Gr\"{o}nniger:         Systemmodell-basierte Definition objektbasierter Modellierungssprachen mit semantischen Variationspunkten\\
2010-18 & Ibrahim Arma\c{c}:         Personalisierte eHomes: Mobilit\"{a}t, Privatsph\"{a}re und Sicherheit\\
2010-19 & Felix Reidl:         Experimental Evaluation of an Independent Set Algorithm\\
2010-20 & Wladimir Fridman, Christof L\"{o}ding, Martin Zimmermann:         Degrees of Lookahead in Context-free Infinite Games\\
2011-01 $^\ast$ &Fachgruppe Informatik:      Jahresbericht 2011\\
2011-02 & Marc Brockschmidt, Carsten Otto, J\"{u}rgen Giesl:         Modular Termination Proofs of Recursive Java Bytecode Programs by Term Rewriting\\
2011-03 & Lars Noschinski, Fabian Emmes, J\"{u}rgen Giesl:         A Dependency Pair Framework for Innermost Complexity Analysis of Term Rewrite Systems\\
2011-04 & Christina Jansen, Jonathan Heinen, Joost-Pieter Katoen, Thomas Noll:         A Local Greibach Normal Form for Hyperedge Replacement Grammars\\
2011-06 & Johannes Lotz, Klaus Leppkes, and Uwe Naumann:         dco/c++ - Derivative Code by Overloading in C++\\
2011-07 & Shahar Maoz, Jan Oliver Ringert, Bernhard Rumpe:         An Operational Semantics for Activity Diagrams using SMV\\
2011-08 & Thomas Str\"{o}der, Fabian Emmes, Peter Schneider-Kamp, J\"{u}rgen Giesl, Carsten Fuhs:         A Linear Operational Semantics for Termination and Complexity Analysis of ISO Prolog\\
2011-09 & Markus Beckers, Johannes Lotz, Viktor Mosenkis, Uwe Naumann (Editors):         Fifth SIAM Workshop on Combinatorial Scientific Computing\\
2011-10 & Markus Beckers, Viktor Mosenkis, Michael Maier, Uwe Naumann:         Adjoint Subgradient Calculation for McCormick Relaxations\\
2011-11 & Nils Jansen, Erika \'{A}brah\'{a}m, Jens Katelaan, Ralf Wimmer, Joost-Pieter Katoen, Bernd Becker:         Hierarchical Counterexamples for Discrete-Time Markov Chains\\
2011-12 & Ingo Felscher, Wolfgang Thomas:         On Compositional Failure Detection in Structured Transition Systems\\
2011-13 & Michael F\"{o}rster, Uwe Naumann, Jean Utke:         Toward Adjoint OpenMP\\
2011-14 & Daniel Neider, Roman Rabinovich, Martin Zimmermann:         Solving Muller Games via Safety Games\\
2011-16 & Niloofar Safiran, Uwe Naumann:         Toward Adjoint OpenFOAM\\
2011-17 & Carsten Fuhs:         SAT Encodings: From Constraint-Based Termination Analysis to Circuit Synthesis
\\
2011-18 & Kamal Barakat:         Introducing Timers to pi-Calculus\\
2011-19 & Marc Brockschmidt, Thomas Str\"{o}der, Carsten Otto, J\"{u}rgen Giesl:         Automated Detection of Non-Termination and NullPointerExceptions for Java Bytecode\\
2011-24 & Callum Corbett, Uwe Naumann, Alexander Mitsos:         Demonstration of a Branch-and-Bound Algorithm for Global Optimization using McCormick Relaxations\\
2011-25 & Callum Corbett, Michael Maier, Markus Beckers, Uwe Naumann, Amin Ghobeity, Alexander Mitsos:         Compiler-Generated Subgradient Code for McCormick Relaxations\\
2011-26 & Hongfei Fu:         The Complexity of Deciding a Behavioural Pseudometric on Probabilistic Automata\\
2012-01 & Fachgruppe Informatik:         Annual Report 2012\\
2012-02 & Thomas Heer:         Controlling Development Processes\\
2012-03 & Arne Haber, Jan Oliver Ringert, Bernhard Rumpe:         MontiArc - Architectural Modeling of Interactive Distributed and Cyber-Physical Systems\\
2012-04 & Marcus Gelderie:         Strategy Machines and their Complexity\\
2012-05 & Thomas Str\"{o}der, Fabian Emmes, J\"{u}rgen Giesl, Peter Schneider-Kamp, and Carsten Fuhs:         Automated Complexity Analysis for Prolog by Term Rewriting\\
2012-06 & Marc Brockschmidt, Richard Musiol, Carsten Otto, J\"{u}rgen Giesl:         Automated Termination Proofs for Java Programs with Cyclic Data\\
2012-07 & Andr\'{e} Egners, Bj\"{o}rn Marschollek, and Ulrike Meyer:         Hackers in Your Pocket: A Survey of Smartphone Security Across Platforms\\
2012-08 & Hongfei Fu:         Computing Game Metrics on Markov Decision Processes\\
2012-09 & Dennis Guck, Tingting Han, Joost-Pieter Katoen, and Martin R. Neuh\"{a}u\ss{}er:         Quantitative Timed Analysis of Interactive Markov Chains\\
2012-10 & Uwe Naumann and Johannes Lotz:         Algorithmic Differentiation of Numerical Methods: Tangent-Linear and Adjoint Direct Solvers for Systems of Linear Equations\\
2012-12 & J\"{u}rgen Giesl, Thomas Str\"{o}der, Peter Schneider-Kamp, Fabian Emmes, and Carsten Fuhs:         Symbolic Evaluation Graphs and Term Rewriting --- A General Methodology for Analyzing Logic Programs\\
2012-15 & Uwe Naumann, Johannes Lotz, Klaus Leppkes, and Markus Towara:         Algorithmic Differentiation of Numerical Methods: Tangent-Linear and Adjoint Solvers for Systems of Nonlinear Equations\\
2012-16 & Georg Neugebauer and Ulrike Meyer:         SMC-MuSe: A Framework for Secure Multi-Party Computation on MultiSets\\
2013-01 $^\ast$ &Fachgruppe Informatik:      Annual Report 2013\\
2013-02 & Michael Reke:         Modellbasierte Entwicklung automobiler Steuerungssysteme in Klein- und mittelst\"{a}ndischen Unternehmen\\
2013-03 & Markus Towara and Uwe Naumann:         A Discrete Adjoint Model for OpenFOAM\\
2013-04 & Max Sagebaum, Nicolas R. Gauger, Uwe Naumann, Johannes Lotz, and Klaus Leppkes:         Algorithmic Differentiation of a Complex C++ Code with Underlying Libraries\\
2013-05 & Andreas Rausch and Marc Sihling:         Software \& Systems Engineering Essentials 2013\\
2013-06 & Marc Brockschmidt, Byron Cook, and Carsten Fuhs:         Better termination proving through cooperation\\
2013-07 & Andr\'{e} Stollenwerk:         Ein modellbasiertes Sicherheitskonzept f\"{u}r die extrakorporale Lungenunterst\"{u}tzung\\
2013-08 & Sebastian Junges, Ulrich Loup, Florian Corzilius and Erika \'{A}brah\'{a}m:         On Gr\"{o}bner Bases in the Context of Satisfiability-Modulo-Theories Solving over the Real Numbers\\
2013-10 & Joost-Pieter Katoen, Thomas Noll, Thomas Santen, Dirk Seifert, and Hao Wu:         Performance Analysis of Computing Servers using Stochastic Petri Nets and Markov Automata\\
2013-12 & Marc Brockschmidt, Fabian Emmes, Stephan Falke, Carsten Fuhs, and J\"{u}rgen Giesl:         Alternating Runtime and Size Complexity Analysis of Integer Programs\\

\end{longtable}
\bigskip

\noindent
{\small $^\ast$ These reports are only available as a printed version.\\
  Please contact \email{biblio@informatik.rwth-aachen.de} to obtain
  copies.}

\end{document}